\documentclass[12pt]{article}
\usepackage{bm}
\usepackage{epsfig}
\usepackage{amssymb}
\usepackage{amsmath}
\usepackage{calrsfs}
\usepackage{multirow}
\usepackage{rotating}
\usepackage{xcolor}
\begin{document}

\setlength{\arraycolsep}{0.5mm}

\title{
SUSY-like relation in evolution of gluon and quark jet multiplicities
\author{Bernd~A.~Kniehl$^{1}$, Anatoly~V.~Kotikov$^{2}$\\
 $^{1}$ {II.} Institut f\"ur Theoretische Physik,\\
Universit\"at Hamburg, Luruper Chaussee 149, 22761 Hamburg, Germany\\
$^{2}$ Bogoliubov Laboratory of Theoretical Physics,\\
Joint Institute for Nuclear Research, 141980 Dubna, Russia
}}

\maketitle

\begin{abstract}
We show
the new
relationship \cite{Kniehl:2017fix} between the anomalous
dimensions, resummed through next-to-next-to-leading-logarithmic order, in the
Dokshitzer-Gribov-Lipatov-Altarelli-Parisi (DGLAP) evolution equations for the
first Mellin moments $D_{q,g}(\mu^2)$ of the quark and gluon fragmentation
functions, which correspond to the average hadron multiplicities in jets
initiated by quarks and gluons, respectively.
This relationship, which is independent of the number of quark flavors, strongly
improves previous treatments by allowing for an exact solution of
the evolution equations.
So far, such relationships have only been known from supersymmetric QCD.
\end{abstract}



The inclusive
production of single hadrons involves the notion of fragmentation functions
$D_a(x,\mu^2)$  (hereafter $(a=q,g)$), where $\mu$ is the factorization scale.
Their $\mu^2$
dependences
are governed by the timelike DGLAP
equations \cite{Gribov:1972ri,Dokshitzer:1977sg}.
The scaling violations, {\it i.e.}, the $\mu^2$ dependences, of $D_a(x,\mu^2)$
may be exploited in global data fits to extract the strong-coupling constant
$\alpha_s=g_s^2/(4\pi)$, leading to very competitive results
\cite{Kniehl:2000fe} as for the world average \cite{Olive:2016xmw}.

The DGLAP equations are conveniently solved in Mellin space, where
$D_a(N,\mu^2)=\int dx\,x^{N-1}D_a(x,\mu^2)$ 
$(N=1,2,\ldots)$, 
because convolutions are converted to products
\begin{equation}
\frac{\mu^2d}{d\mu^2}
\left(\begin{array}{l} D_s(N,\mu^2) \\ D_g(N,\mu^2) \end{array}\right)
=\left(\begin{array}{ll} P_{qq}(N) & P_{gq}(N) \\
P_{qg}(N) & P_{gg}(N) \end{array}\right)
\left(\begin{array}{l} D_s(N,\mu^2) \\ D_g(N,\mu^2) \end{array}\right),
\label{apR}
\end{equation}
where $P_{ab}(N)$ 
$(a,b=q,g)$
are anomalous dimsnsions and
$D_s=(1/2n_f)\sum_{q=1}^{n_f}(D_q+D_{\bar{q}})$, with $n_f$ being the
number of active quark flavors, is the quark singlet component.
The quark non-singlet component
is irrelevant for the present study.

The first Mellin moment $D_a(\mu^2)\equiv D_a(1,\mu^2)$ is of particular
interest in its own right because, up to corrections of orders beyond our
consideration here, it corresponds to the average hadron multiplicity
$\langle n_h\rangle_a$ of jets initiated by parton $a$.
There exists a wealth of experimental data on $\langle n_h\rangle_q$,
$\langle n_h\rangle_g$, and their ratio
$r=\langle n_h\rangle_g/\langle n_h\rangle_q$ for charged hadrons $h$ taken in
$e^+e^-$ annihilation at various center-of-mass energies $\sqrt{s}$, ranging
from 10 to 209~GeV  (see
\cite{Bolzoni:2013rsa}).
The study of $D_a$ is a topic of old vintage; the LO value of $r$,
$C^{-1}=C_A/C_F$ with color factors $C_F=4/3$ and $C_A=3$, was found four
decades ago \cite{Brodsky:1976mg}.

The description of the $\mu^2$ dependences of $D_a$ at fixed order in
perturbation theory are spoiled by the fact that $P_{ba}\equiv P_{ba}(1)$ are
ill defined and require resummation, which was performed for the leading
logarithms (LL) \cite{Mueller:1981ex}, the next-to-leading logarithms (NLL)
\cite{Vogt:2011jv}, and the next-to-next-to-leading logarithms (NNLL)
\cite{Kom:2012hd}.

In Ref. \cite{Kniehl:2017fix},
we exposed a relationship between the NNLL-resummed expressions
for $P_{ba}$, which has gone unnoticed so far.
Its existence in QCD is quite remarkable and interesting in its own right,
because a similar relationship is familiar from supersymmetric (SUSY) QCD,
where $C=1$
\cite{Dokshitzer:1977sg,Kom:2012hd}.
Owing to this new relationship, the DGLAP equations may be solved exactly,
which greatly consolidates the theoretical foundation for the determination of
$\alpha_s$ and thus reduces its theoretical uncertainty.

Our starting point is Eq.~(\ref{apR}) for $N=1$ with NNLL resummation.
We have \cite{Kom:2012hd}
\begin{eqnarray}
P_{aa}&=&\gamma_0(\delta_{ag} + K_{a}^{(1)} \gamma_0 
+ K_{a}^{(2)}  \gamma_0^2) ,~~
P_{gq}=
C (P_{gg} +A),~~
P_{qg} =
C^{-1} (P_{qq} +A) ,
\label{NNLL}
\end{eqnarray}
with $\mathcal{O}(\gamma_0^3)$ accuracy,
where $\gamma_0=\sqrt{2C_Aa_s}$, 
$a_s=\alpha_s/(4\pi)$ is
the couplant,
$\delta_{ab}$ is the Kronecker symbol, and
\begin{eqnarray}
K_{q}^{(1)} &=& \frac{2}{3} C\varphi,\quad K_{g}^{(1)} =
-\frac{1}{12}[11 +2\varphi (1+6C)],\quad
K_{q}^{(2)} = -\frac{1}{6} C\varphi [17-2\varphi(1-2C)],
\nonumber \\
K_{g}^{(2)} &=& \frac{1193}{288} -2\zeta(2)
- \frac{5\varphi}{72}(7-38C)+\frac{\varphi^2}{72}(1-2C)(1-18C),~
A =
K_{q}^{(1)}\gamma_0^2,~
\varphi=\frac{n_f}{C_A}.
\label{nllfirstA}
\end{eqnarray}
Eq.~(\ref{NNLL}) is written in a form that allows us to glean a novel
relationship:
\begin{equation}
C^{-1}P_{gq}-P_{gg}=CP_{qg}-P_{qq},
\label{Basic}
\end{equation}
which is independent of $n_f$. It
generalizes the case of SUSY QCD
\cite{Dokshitzer:1977sg,Kom:2012hd} where
$C=1$.
            
We now solve Eq.~(\ref{apR}) exactly at $N=1$ by exploiting Eq.~(\ref{Basic}).
To this end, we diagonalize the NNLL DGLAP evolution kernel as
\begin{equation}
U^{-1}\left(\begin{array}{ll}
P_{qq} & P_{gq} \\ P_{qg} & P_{gg}
\end{array}\right)U
=\left(\begin{array}{ll}
P_{--} & 0 \\ 0 & P_{++}
\end{array}\right),\quad
U=\left(\begin{array}{ll}
1 & -1 \\ \frac{1-\alpha}{\varepsilon} & \frac{\alpha}{\varepsilon}
\end{array}\right),\quad
U^{-1}=\left(\begin{array}{ll} 
\alpha & \varepsilon \\ \alpha-1 & \varepsilon
\end{array}\right),
\label{matrix}
\end{equation}
where 
\begin{eqnarray}
\alpha=
\frac{P_{qq}-P_{++}}{P_{--}-P_{++}},~
\varepsilon=\frac{P_{gq}}{P_{--}-P_{++}},~
P_{\pm\pm}=
\frac{1}{2}\left[P_{qq}+P_{gg}\pm
\sqrt{(P_{qq}-P_{gg})^2+4P_{qg}P_{gq}}\right].\qquad
\label{Ppm}
\end{eqnarray}
Eq.~(\ref{apR}) thus assumes the form
\begin{equation}
\frac{\mu^2d}{d\mu^2}\left(\begin{array}{l} D_- \\ D_+ \end{array}\right)
=\left[
\left(\begin{array}{ll} P_{--} & 0 \\ 0 & P_{++}\end{array}\right)
-U^{-1}\frac{\mu^2d}{d\mu^2}U\right]
\left(\begin{array}{l} D_- \\ D_+ \end{array}\right),
\label{ap2a}
\end{equation}
where the second term contained within the square brackets stems from the
commutator of $\mu^2d/d\mu^2$ and $U$, and
\begin{equation}
\left(\begin{array}{l} D_- \\ D_+ \end{array}\right)
=U^{-1}\left(\begin{array}{l} D_s \\ D_g \end{array}\right)
=\left(\begin{array}{l}\alpha D_s+\varepsilon D_g \\
(\alpha-1)D_s+\varepsilon D_g \end{array}\right).
\label{ap1.2}
\end{equation}
Owing to Eq.~(\ref{Basic}), the square root in
Eq.~(\ref{Ppm})  is exactly canceled,
and
we have simple expressions for $P_{\pm \pm}$
\begin{eqnarray}
P_{--}=
-A,\qquad
P_{++}=P_{qq}+P_{gg}+A,\qquad
\alpha=
\frac{P_{gg}+A}{P_{qq}+P_{gg}+2A},\qquad
\varepsilon = -C \alpha \, .
\label{alpha1}
\end{eqnarray}
Inserting two last equalities
of Eq.~(\ref{alpha1}) in Eq.~(\ref{matrix}), we
have
\begin{equation}
U^{-1}\frac{\mu^2d}{d\mu^2}U=-\frac{1}{\alpha}\,\frac{\mu^2d}{d\mu^2}\alpha
\left(\begin{array}{ll} 1 & 0 \\ 1 & 0\end{array}\right),~~
\frac{1}{\alpha}\,\frac{\mu^2d}{d\mu^2}\alpha
=\frac{C\varphi\beta_0}{3C_A}\gamma_0^3+\mathcal{O}(\gamma_0^4),~~
\beta_0=\frac{C_A}{3}(11-2\varphi) .
\label{eq:da}
\end{equation}
Inserting Eq.
(\ref{eq:da}) in
Eq.~(\ref{ap2a}), we may cast Eq.~(\ref{apR}) in its final form,
\begin{equation}
\frac{\mu^2d}{d\mu^2}
\left(\begin{array}{l} D_- \\ D_+ \end{array}\right)
= \left(\begin{array}{ll} \frac{C\varphi\beta_0}{3C_A}\gamma_0^3-A & 0 \\ 
\frac{C\varphi\beta_0}{3C_A}\gamma_0^3 & P_{gg}+P_{qq} + A \end{array}\right) 
\left(\begin{array}{l} D_- \\ D_+ \end{array}\right).\quad
\label{ap2b}
\end{equation}
The initial conditions are given by Eq.~(\ref{ap1.2}) for $\mu=\mu_0$ in terms
of the three constants $\alpha_s(\mu_0^2)$, $D_s(\mu_0^2)$, and $D_g(\mu_0^2)$.

The solution of Eq.~(\ref{ap2b}) is greatly facilitated by the fact that one
entry of the matrix on its right-hand side is zero.
We may thus obtain $D_-$  exactly
\begin{equation}
\frac{D_-(\mu^2)}{D_-(\mu_0^2)} =
\exp\!{\left[\int_{\mu_0^2}^{\mu^2}\!\!\!\frac{d\bar{\mu}^2}{\bar{\mu}^2}
\!\left(\frac{C\varphi\beta_0}{3C_A}\gamma_0^3-A\!\right)\!\right]}
= \frac{T_-(\gamma_0(\mu^2))}{T_-(\gamma_0(\mu_0^2))}, ~~
T_-(\gamma_0)=
\gamma_0^{d_-}\exp{\left(-\frac{4}{3}C\varphi \gamma_0\right)},
\label{gensol.-}
\end{equation}
with $d_-=8C_AC\varphi/(3\beta_0)$.
The 
correction $\propto\gamma_0$ in $T_-(\gamma_0)$
originates from
the extra term in Eq.~(\ref{ap2a}) and represents a novel feature of our
approach.
In Ref.~\cite{Bolzoni:2013rsa} and analogous analyses for
parton densities
\cite{Kotikov:1998qt}, the minus
components do not participate in the resummation.

We are then left with an inhomogeneous differential equation for $D_+$.
The general solution $\tilde{D}_+$ of its homogeneous part reads
\begin{eqnarray}
&&\frac{\tilde{D}_+(\mu^2)}{\tilde{D}_+(\mu_0^2)}=
\exp{\left[\int_{\mu_0^2}^{\mu^2}\frac{d\bar{\mu}^2}{\bar{\mu}^2}
\gamma_0\Bigl(1+ (2K_{q}^{(1)}+K_{g}^{(1)})
\gamma_0+(K_{q}^{(2)}+K_{g}^{(2)})
\gamma_0^2\Bigr)\right]}
=\frac{T_+(\gamma_0(\mu^2))}{T_+(\gamma_0(\mu_0^2))}, \nonumber \\
&&T_+(\gamma_0)
=\gamma_0^{d_+}\exp{\left[\frac{4C_A}{\beta_0\gamma_0}
-\frac{4C_A}{\beta_0} \left(K_{q}^{(2)}+K_{g}^{(2)}
-b_1\right)\gamma_0\right]},~~\beta_1=\frac{2C_A^2}{3}[17-\varphi(5+3C)],
\label{gensol.+}
\end{eqnarray}
where
$d_+=-4C_A(2K_{q}^{(1)}+K_{g}^{(1)})/
\beta_0$ and $b_1=\beta_1/(2C_A \beta_0)$.
Adding to $\tilde{D}_+$ a special solution of the inhomogeneous differential
equation for $D_+$, we find its general solution to be 
\begin{eqnarray}
D_+(\mu^2)=
\left[\frac{D_+(\mu_0^2)}{T_+(\gamma_0(\mu_0^2))}
-\frac{4}{3}C\varphi\frac{D_-(\mu_0^2)}{T_-(\gamma_0(\mu_0^2))}
\int_{\gamma_0(\mu_0^2)}^{\gamma_0(\mu^2)}
\frac{d\gamma_0}{1+b_1\gamma_0^2}\,\frac{T_-(\gamma_0)}{T_+(\gamma_0)}\right]
T_+(\gamma_0(\mu^2)).
\label{gensol.+a}
\end{eqnarray}
The final expressions for $D_-$ and $D_+$ in Eqs.~(\ref{gensol.-}) and
(\ref{gensol.+a}), respectively, are fully renormalization group improved
because all $\mu$ dependence resides in $\gamma_0$.

The results (\ref{gensol.-}) and (\ref{gensol.+a}) allow us
to perform a global fit to the available measurements
of $\langle n_h\rangle_q$ and $\langle n_h\rangle_g$ for changed hadrons $h$ in
$e^+e^-$ annihilation, which can be found in
\cite{Kniehl:2017fix,Bolzoni:2013rsa}.
Referring the interested reader to Ref. \cite{Kniehl:2017fix}, we notice only that we got at
68\% CL, 
   \begin{equation}
\alpha_s^{(5)}(M_Z^2)=0.1205\genfrac{}{}{0pt}{}{+0.0016}{-0.0020},
\label{eq:as}
\end{equation}
which nicely agrees with the present world average,
$\alpha_s^{(5)}(M_Z^2)=0.1181\pm0.0011$ \cite{Olive:2016xmw}.

In summary, we shown
an unexpected, SUSY-like relationship \cite{Kniehl:2017fix} between the
NNLL-resummed first Mellin moments of the timelike DGLAP splitting functions
in real QCD, Eq.~(\ref{Basic}), which is $n_f$ independent, and exploited it to
find an exact solution of the DGLAP evolution equation, Eq.~(\ref{apR}) at $N=1$.

This research was supported in part 
by the German Research Foundation under Grant No.\ KN~365/5-3,
by the Russian Foundation for Basic Research under Grant No.\ 16-02-00790-a,
and by the Heisenberg-Landau Programme.

\end{document}